\documentclass{kapproc} % Computer Modern font calls

\usepackage{procps}

\usepackage[dvips]{graphicx}

\usepackage{epsfig}

\useuppercase

\normallatexbib

\begin{document}
%\flushright
\articletitle{Direct measurements of tunable Josephson plasma
resonance in the L-SET} % tää sopiiki hyvin teemaan että tuunataan kahella knobilla :-)

\author{Mika A. Sillanpää, Leif Roschier, and Pertti J. Hakonen}
\affil{Low Temperature Laboratory, Helsinki University of
Technology\\
Otakaari 3 A, Espoo P.O.Box 2200 FIN-02015 HUT Finland}
%\email{Mika.Sillanpaa@iki.fi}

\begin{abstract}
Phase dynamics has been measured in a driven mesoscopic Josephson
oscillator where the resonance is tuned either by magnetic flux or
by gate charge modulation of the Josephson inductance. Phenomena
are analyzed in terms of a "phase particle picture", and by
numerical circuit simulations. An analogy to switching of a
DC-biased junction into voltage state is discussed. Operation
principle of the recently demonstrated Inductive Single-Electron
Transistor (L-SET) based on the driven oscillator is reviewed. The
obtained charge sensitivity implies that a performance comparable
to the best rf-SETs has already been reached with the novel
device.
\end{abstract}

\section{Introduction}

Quantum measurement in the solid state has been shown to be
feasible as several impressive realizations of qubits based on
mesosocopic superconducting tunnel junctions have emerged
\cite{nakamuraqb,vion,hanqb,martinisqb}. Sensitive measurement of
physical quantities close to the limit set by the uncertainty
principle is, on the other hand, an important issue in its own
right.
% Due to typically high amount of $1/f$ noise in mesoscopic
%systems, an input bandwidth extending beyond the $1/f$ corner is
%necessary as well in addition to high measurement sensitivity.

The Single-Electron Transistor (SET) is a basic mesoscopic
detector, sensitive to electric charge on a gate capacitor. In
order to gain advantage of the inherently large bandwidth
$(R_{SET} C_{\Sigma})^{-1} \sim$ 10 GHz of the SET charge
detector, basically two new technologies have been developed where
the SET is read using an $LC$ oscillator built from "macroscopic"
components but coupled directly to it. The "rf-SET"
(Radio-Frequency SET) \cite{rfset} is based on gate modulation of
the $Q$-value of the oscillator.

Because of limitations due to the dissipative nature of the
rf-SET, a principally non-dissipative "L-SET" (Inductive SET)
technique has been developed very recently \cite{lset}. The L-SET
is based on reactive readout of the Josephson inductance of a
superconducting SET (SSET) in a resonator configuration
\cite{zorinrf}. Due to correlated Cooper pair tunneling, it does
not exhibit shot noise or excessive dissipation.

In this paper we first review the operating principle of the L-SET
charge detector. Then we concentrate on discussing classical
dynamics of the phase $\varphi$ under the microwave drive, and we
present new experimental data. In particular, we discuss dynamical
effects in the L-SET resonator which resemble the switching of a
DC-biased Josephson junction into voltage state by drive or noise.

\section{The L-SET circuit}

\begin{figure}[ht]
\centering \epsfig{figure=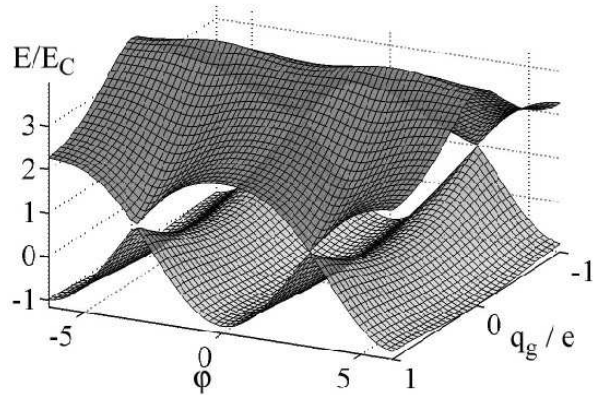, width=8cm} \vskip.2in
\caption{Two lowest bands $n = 0$ and $n = 1$ from Eq.\ \ref{eq:H}
for $E_J/E_C = 1.7$ (sample 2) plotted as function of the gate
charge $q_g$ and the phase $\varphi$ across the SSET.}
\label{fig:band}
\end{figure}

In the absence of DC bias voltage, a SSET has the Hamiltonian

\begin{equation}\label{eq:H}
H = \frac{(q-q_g)^2}{2 C_{\Sigma}} - 2 E_J \cos(\varphi/2)
\cos(\theta),
\end{equation}

\noindent where $q_g$ is the gate charge, and $\varphi$ is the
phase difference across the SSET, assumed to be a classical
variable here due to an environment having impedance much smaller
than $R_Q \simeq 6.5$ k$\Omega$. $E_J$ is the single-junction
Josephson energy, and the charging energy is related to sum
capacitance by $E_C = e^2/(2 C_{\Sigma})$. Eigenvalues of this
well-known Hamiltonian form bands $E_n (\varphi, q_g)$ (see Fig.\
\ref{fig:band}).

At the lowest band $n = 0$, the energy $E_0$ increases
approximately quadratically as a function of the phase $\varphi$
when moved to either direction from the minimum at $\varphi = 0$.
This type of dependence is characteristic of an inductor. The
effective Josephson inductance of the SSET is then

\begin{equation} \label{eq:lj}
L_J^{*} = \frac{\Phi_0 ^2}{2 \pi} (E_J^{*})^{-1},
\end{equation}

\noindent where the Josephson coupling has the effective value

\begin{equation} \label{eq:ej}
E_J^{*} = \frac{ \partial ^2 E_0(q_g, \varphi)}{ \partial
\varphi^2}.
\end{equation}

\noindent Here, $\Phi_0 = h/(2 e)$ is the flux quantum.

Since the energy band, and consequently, the inductance, depend on
the gate charge $q_g$, the L-SET electrometer is built so that the
resonance frequency of an $LC$ tank circuit connected to a SSET is
tunable by $q_g$. This allows in principle a purely reactive
readout.

\begin{figure}[ht]
\centering \epsfig{figure=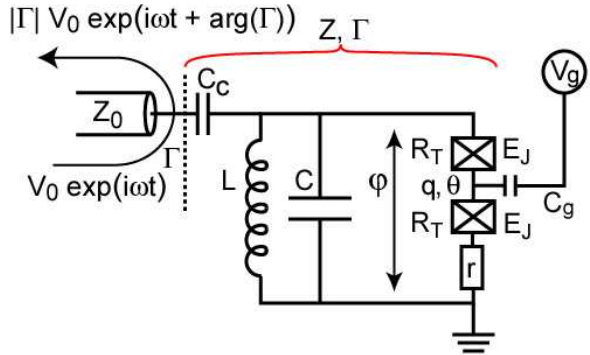, width=8cm} \vskip.2in
\caption{Schematics of the L-SET resonance circuit coupled to
microwave feedline. The resonance is read by measuring amplitude
or phase of the reflection coefficient $\Gamma = (Z-Z_0)/(Z+Z_0)$,
where $Z_0 = 50 \Omega$, and $Z(L_J^{*})$ is the resonator
impedance (including $C_c$).} \label{fig:circuit}
\end{figure}

The L-SET circuit we use is shown in Fig. \ref{fig:circuit} where
the SSET is coupled in parallel to an $LC$ oscillator resonant at
the frequency $f_0 = 1/(2 \pi) (L C)^{-1/2}$, roughly at 600 MHz.
The total system has the gate-dependent plasma resonance at $f_p =
1/(2 \pi) (L_{tot} C)^{-1/2} > f_0$, where $L_{tot} = L \parallel
L_{J}^*$. The bandwidth $\Delta f \simeq f_p / Q_e$, where $Q_e$
is the coupled quality factor, is typically in the range of tens
of MHz.

The resistor $r$ in series with the SSET is a model component for
dissipation. As compared to the more standard way of drawing a
resistor in parallel with the resonator, we got here a better
agreement with the non-linear dynamics (see section
\ref{sec:switch}).

Charge detection is performed by measuring a change of amplitude
or phase of the voltage reflection coefficient $\Gamma$, when the
setup is irradiated by microwaves of frequency roughly $f_p$. In
the best sample so far \cite{tbb}, we have measured charge
sensitivity $3 \times 10^{-5}$e$/\sqrt{\mathrm{Hz}}$ over a
bandwidth of about 100 MHz.

%The Josephson phase $\varphi$ across the SSET in the L-SET
%configuration experiences classical dynamics, interesting in
%itself, whose theoretical considerations we concentrate on next.

\section{Plasma oscillations in L-SET}

Dynamics of the L-SET oscillator can be analyzed in terms of a
potential $E_{p} = E_{n} (\varphi, q_g)  + E_L$ due to the
Josephson inductance and the shunting $L$, respectively, at the
ground band approximately

\begin{equation} \label{eq:pot}
E_{p} = - E_{0} (q_g) \cos(\varphi) + \Phi_0 ^2 /(8 \pi^2 L)
\varphi^2.
\end{equation}

\begin{figure}[ht]
\centering \epsfig{figure=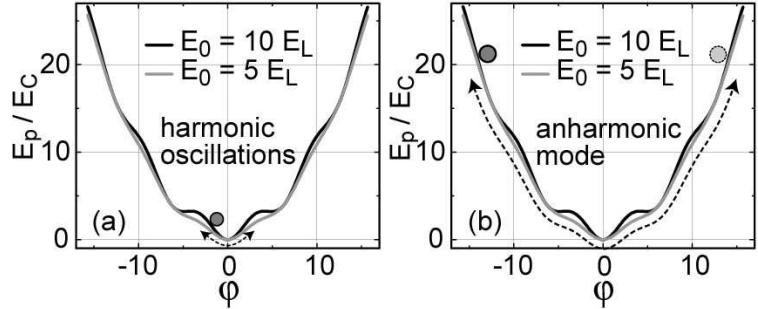, width=10cm}
\vskip.2in \caption{Illustration of the fictitious phase particle
in a sinusoidally modulated parabolic potential.} \label{fig:pot}
\end{figure}

At small oscillation amplitude, the phase particle experiences
harmonic oscillations around $\varphi = 0$, whose frequency $f_p$
is controlled by gate-tuning of the Josephson inductance (Fig.\
\ref{fig:pot} (a)). This mode of operation, where the L-SET works
as a charge-to-frequency converter, is the "harmonic mode".

Once the oscillation amplitude is increased close to the critical
current at $\varphi = \pm \pi/2$, the $\cos(\varphi)$ term changes
the local curvature of the potential, and hence, the resonance
frequency changes. At very high amplitude, Fig.\ \ref{fig:pot}
(b), the cosine wiggle becomes averaged out. Thus, we expect a
change of resonant frequency from $f_p$ to $f_0$ roughly at an AC
current of critical current magnitude. This change of resonance
frequency when the sample is probed by \emph{critical power} $P_c$
is reminiscent of a DC-biased Josephson junction switching into a
voltage state \cite{yale}.

At large excitations above $P_c$, the highly nonlinear oscillator
experiences complicated dynamics which does not in general allow
analytical solutions. Numerical calculations over a large range of
parameters, however, show consistently that the system response
depends on $L_J^*$ also in this case \cite{tbb}. This mode of
operation of the L-SET charge detector we call the "anharmonic"
mode.

\section{Simulation scheme}

\begin{figure}[ht]
\epsfig{figure=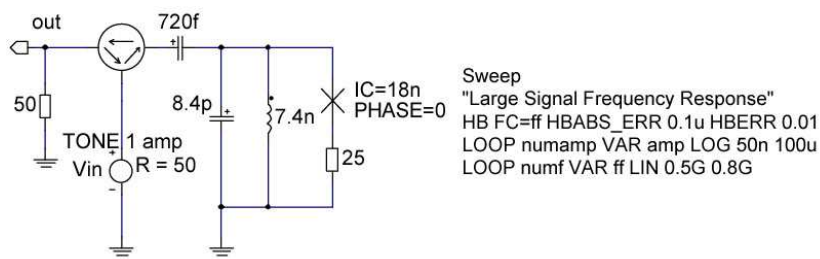, width=11cm} \vskip.2in
\caption{Screenshot from Aplac program used to simulate classical
dynamics in the L-SET circuit (to aid convergence, we also used 1
k$\Omega$ across the SSET).} \label{fig:aplacscreen}
\end{figure}

We simulated the transition from linear to non-linear oscillations
in the L-SET circuit with the Aplac circuit simulation program
which contains an implementation of a Josephson junction element.
The SSET was modeled as a single tunable junction. We used the
method of harmonic balance where amplitudes of the first and three
upper harmonics were optimized to create an approximate solution.
Screenshot of the simulation schematics is shown in Fig.\
\ref{fig:aplacscreen}.

\section{Experiment}

The measurements were done in a powerful Leiden Cryogenics
MiniKelvin dilution refrigerator. At the base temperature of 20
mK, we used a microwave circulator which had about 20 dB backward
isolation to cut back-action noise from the 4 Kelvin preamplifier
\cite{amplifier}.

The reflected probing excitation was amplified with a chain of
amplifiers having a total of 5 K noise temperature, and detected
with a network analyzer or a spectrum analyzer.

\begin{figure}[ht]
\centering \epsfig{figure=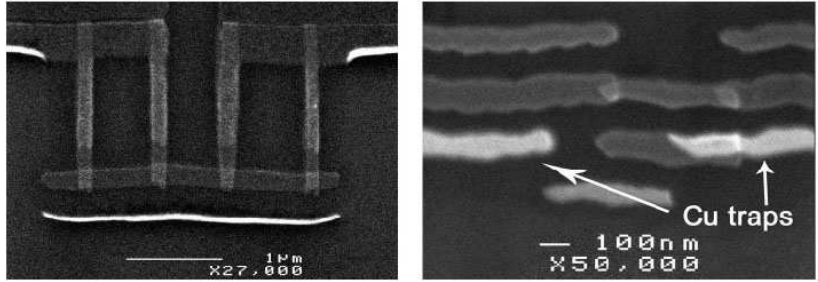, width=11cm} \vskip.2in
\caption{SEM micrographs of the samples discussed; sample 1 with
SQUID junctions (left), and sample 2 (right) which had
quasiparticle traps which were in contact to the electrodes $5
\mu$m from junctions.} \label{fig:semimage}
\end{figure}

We studied two samples (Table \ref{tb:sample} and Fig.\
\ref{fig:semimage}) where sample 1 had a tunable $E_J$. Since it
had $E_C \simeq k_B T$, where the temperature was probably set by
leakage through the circulator, its response had a hardly
detectable gate modulation. Accordingly, we were able to study the
plasma resonance in an almost classical junction.

Sample 2 (discussed in Ref.\ \cite{lset}, see section
\ref{sec:charge}) was a sensitive detector, with a 15 MHz gate
shift of the resonance frequency.

\begin{table}[ht]
\caption{Parameters of the two samples and of their tank $LC$
oscillators as discussed in the text. $E_{J} = h \Delta /(8 e^2
\frac{1}{2} R_{SET})$ is the single-junction Josephson energy, and
$E_C = e^2/(2 C_{\Sigma})$ is the charging energy. For sample 1,
SQUID structure of the individual SET junctions allowed tuning of
the $E_J/E_C$ ratio.} \label{tb:sample}
\begin{tabular*}{\textwidth}{@{\extracolsep{\fill}}lllllllll}
\sphline
sample & $R_{SET}$ (k$\Omega$) & $E_J$ (K) & $E_C$ (K)
&$L_J^{*}$ (nH) & $L$ (nH) & $C$ (pF) & $C_c$ (pF) & $Q_e$ \\
\hline
  1 & 4.2 & 3.5 ... 0 & 0.17 & 6 & 3 & 23 & 0.72 & 13 \\ \hline
  2 & 9.6 & 1.6 & 0.92 & 16 & 7.4 & 8.4 & 0.72 & 18 \\
\sphline
\end{tabular*}
\end{table}

\section{Results and discussion}

\subsection{Charge detection}
\label{sec:charge}

\begin{figure}[ht]
\epsfig{figure=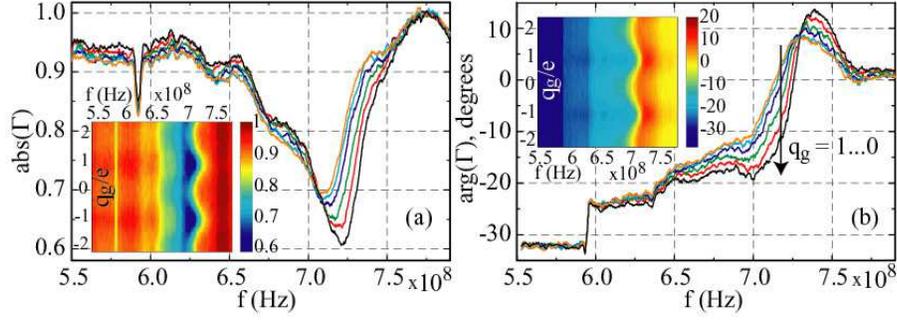, width=12cm} \vskip.2in
\caption{Sample 2; measured frequency response at $-125$ dBm for
successive gate values $q_g/e = 1 \ldots 0$, (a) amplitude, (b)
phase. The insets portray the complete $2e$ periodicity
\cite{lset}.} \label{fig:response}
\end{figure}

For sample 2, $E_C \gg k_B T$, and hence, phase was properly
localized in the potential. Accordingly, we got a rather good
agreement for gate modulation of the resonance frequency (Fig.\
\ref{fig:response}), where the resonance moves 15 MHz with respect
to gate charge. This agrees well with the expectation of only 15
\% modulation of $L_J^*$ based on the rather high $E_J / E_C
\simeq 1.7$ of this sample. The minimum (w.r.t. gate) of $L_J^*$,
fitted best with 18 nH is about 20 \% higher than expected. This
is probably due to a small amount of phase noise, or partially due
to inaccuracies in determination of the parameters.

Charge sensitivity was measured using amplitude readout. In the
harmonic mode we got $s_q = 2.0 \times
10^{-3}$e$/\sqrt{\mathrm{Hz}}$  at the maximum power $P_c \simeq
-116$ dBm, corresponding to 20 fW dissipation in the whole
resonator circuit. Note that due to Cooper pair tunneling, the
power is not dissipated in the SSET island. In the anharmonic
mode, significantly better sensitivity of $s_q = 1.4 \times
10^{-4}$e$/\sqrt{\mathrm{Hz}}$ was obtained.

In the best sample so far \cite{tbb}, having $E_J / E_C \simeq
0.3$ we have measured $3 \times 10^{-5}$e$/\sqrt{\mathrm{Hz}}$
over a bandwidth of about 100 MHz. This kind of performance is
comparable to the best rf-SET's \cite{chalmers}, though power
dissipation is more than two orders of magnitude lower.

\subsection{Harmonic oscillations}

\begin{figure}[ht]
\epsfig{figure=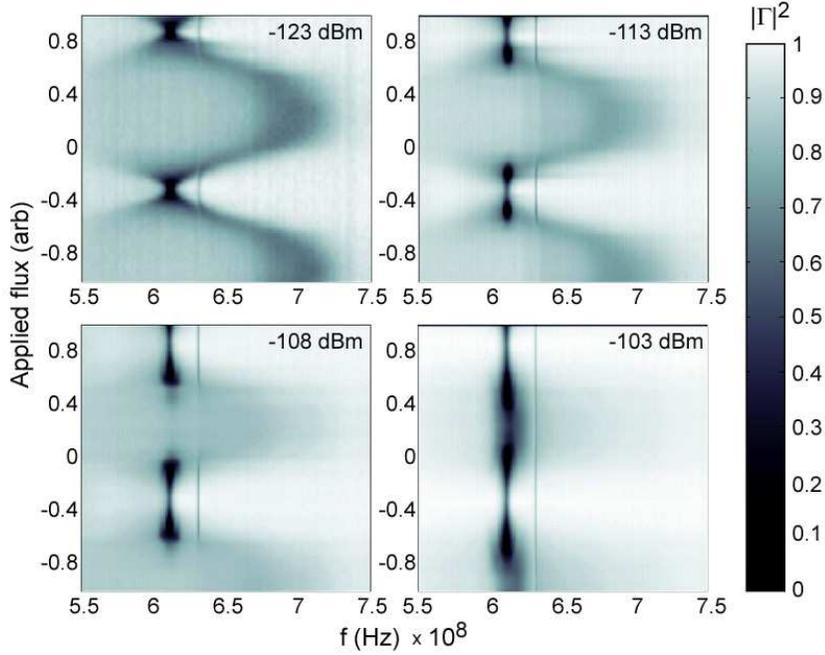, width=11cm} \vskip.2in
\caption{Flux modulation of the power reflection ($ | \Gamma |^2
$) for sample 1 in order of increasing probing excitation (the
graph labelled -103 dBm corresponds approximately to critical
power $P_c$ at maximum $E_J$).} \label{fig:harm}
\end{figure}

In Fig.\ \ref{fig:harm} we plot data from sample 1 which behaves
almost as a classical junction with negligible charging effect. By
applied magnetic flux we tune simultaneously $E_J$ of both
SQUID-shaped junctions of the SSET, and hence, $L_J^*$.
Periodicity of the resonance frequency with respect to applied
magnetic flux is evident.

A prominent feature in Fig.\ \ref{fig:harm} is the absence of
reflected signal at "critical" points where the resonance
frequency switches to $f_0$. This implies a strong dissipation
which is visible as coupling to $Z_0$. Pronounced dissipation at
the critical points is representative of strong fluctuations at
the artificial "phase transition" points.

\begin{figure}[ht]
\centering \epsfig{figure=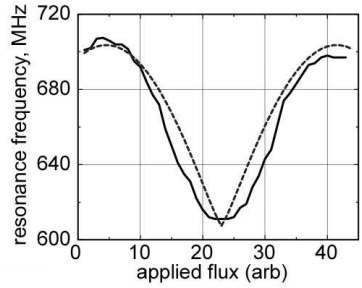, width=5cm} \vskip.2in
\caption{Measured resonance frequency for sample 1 extracted from
the first graph in Fig.\ \ref{fig:harm} (-123 dBm). The dotted
line is a fit based on bare flux modulation of the $E_J/E_C$
ratio.} \label{fig:resfreq}
\end{figure}

In Fig.\ \ref{fig:resfreq} we plot $f_p$ measured at a low
excitation, roughly 1/10 of critical current peak amplitude. The
data are fitted to theoretical flux $\Phi$ modulation of $L_J^*$.
A symmetric SQUID has been assumed, where $E_J \propto \cos (\pi
\Phi / \Phi_0)$ which affects the $E_J/E_C$ ratio and thus $L_J^*$
(see Eqs.\ \ref{eq:lj} and \ref{eq:ej}). It is clear that
something else is happening at low $E_J$, where the resonance
meets $f_0$ more rapidly than expected. Rounding of the cusp in
experimental data cannot be explained by asymmetry in the SQUIDS
either, since then the resonance would deviate from true $f_0$
which is not the case.

We argue this effect is due to "premature switching" caused by
noise in the oscillator. The effect is again in analogy to physics
in a DC-biased junction, namely, noise-induced switching and
delocalization of phase. In contrast to running of phase in a
tilted washboard, the average of phase stays at zero in our case.
Noise-induced switching happens when peak phase fluctuations reach
$\varphi_c \simeq \pi / 2$. This condition corresponds to random
motion of the phase particle at the bottom of the potential of
Fig.\ \ref{fig:pot} (a) with rms amplitude $\sim \varphi_c / 2.5$.

Noise excites the phase particle in the potential $E_{p}$ given by
Eq.\ \ref{eq:pot} up to an energy $E_{p} = k_B T_{eff}$, where
$T_{eff}$ is the electron temperature set by noise. Thus,
approximately $k_B T_{eff} = \Phi_0 ^2 (\varphi_c / 2.5) ^2/(8
\pi^2 L_{tot})$. Since in Fig.\ \ref{fig:resfreq} the switching
happens at the flux $\simeq 0.8 \times \pi \Phi_0 / 2$, we get an
estimate $T_{eff} \simeq 0.5$ K which is likely caused by leakage
outside the band of the circulator.

\subsection{Switching and nonlinear oscillations}
\label{sec:switch}

\begin{figure}[ht]
\epsfig{figure=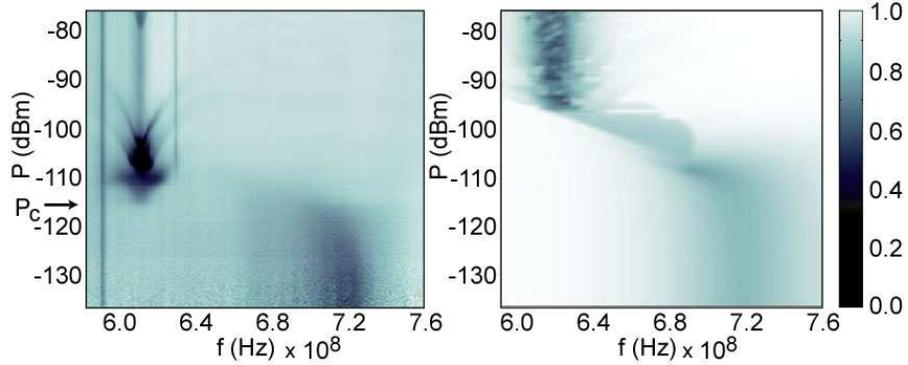, width=12cm} \vskip.2in
\caption{Measured frequency response ($ | \Gamma | $, left) for
sample 2 as a function of excitation power. "Critical power" $P_c$
(see text) is marked. Gate DC level was tuned approximately to the
minimum of $L_J^*$, i.e., the maximum of $f_p$. On the right is
shown a corresponding Aplac simulation run using the circuit of
Fig.\ \ref{fig:aplacscreen}.} \label{fig:freqpower}
\end{figure}

In this section, the discussion is based on experimental data from
sample 2. In Fig.\ \ref{fig:freqpower}, at $P_c \simeq -116$ dBm,
the resonant frequency switches from the broad plasma resonance,
centered at 720 MHz, into a narrower tank resonance at 613 MHz.
The wavelike texture at $-105 \ldots -90$ dBm is due to the $\cos
( \varphi )$ Josephson-potential.

Changes in coupling, e.g., that $Z$ goes through critical coupling
at $-105$ dBm, and sharpening of the resonance above $P_c$, are
due to an increase of the internal $Q$-value $Q_i$ from 20 up to
several hundreds. Since $Q_i \simeq \omega_0 L_J^* / r$, and since
$L_J^*$ acquires a large effective value at a high drive $\gg 2
\pi$ due to cancelling of positive and negative contributions, the
supposed dissipation of $r$ in series with the SSET has less
effect at $P \gg P_c$.

The source of the dissipation modelled by the resistor $r$,
probably situated in the SSET itself, is presently unknown, but it
may be related to quasiparticle poisoning. It also limits the
quality factor up to 20 approximately.

Also plotted in Fig.\ \ref{fig:freqpower} is an Aplac simulation
for the circuit. It presents a qualitative agreement with
experiment and predicts roughly correctly the end of the linear
regime of plasma resonance. However, in experiment the switching
is markedly sharp, which we attribute to the effect of higher
bands of the SSET \cite{moriond}.

\begin{acknowledgments}
It is a pleasure to thank T. Heikkilä, G. Johansson, R. Lindell,
H. Seppä, and J. Viljas for collaboration. This work was supported
by the Academy of Finland and by the Large Scale Installation
Program ULTI-3 of the European Union (Contract
HPRI-1999-CT-00050).
\end{acknowledgments}

%\section*{References}

\end{document}